# Awakening Sleeping Beauties from articles on mRNA vaccines against COVID-19

Artemis Chaleplioglou, Efstathia Selinopoulou, Konstantinos Kyprianos, Alexandros Koulouris

Department of Archival, Library & Information Studies, University of West Attica, 12243 Athens, Greece

e-mails:     Artemis Chaleplioglou, artemischal@uniwa.gr

Efstathia Selinopoulou, mslam236682011@uniwa.gr

Konstantinos Kyprianos, kkyprian@uniwa.gr

Alexandros Koulouris, akoul@uniwa.gr

**Abstract:**  The COVID-19 outbreak rapidly became a pandemic in the first quarter of 2020, posing an unprecedented threat and challenge to healthcare systems and the public. Governments in nearly every country focused on immunization programs for the general population using mRNA vaccines against this disease, marking the first large-scale use of this technology. Previously overlooked research papers on mRNA vaccine preparation or administration gained prominence. The impact was documented bibliographically through a surge in citations these papers received. These reports exemplify the "Sleeping Beauty" bibliometric phenomenon, while the articles that triggered this awakening act as the "Sweet Prince," leading to the resurgence of the previous papers' bibliometric impact. Here, a backward reference search was performed in the Scopus bibliographic database to identify "Sleeping Beauties" by applying the Beauty Coefficient metric. A total of 915 original research articles were published in 2020, citing 21,979 referenced papers, including 1,181 focused on mRNA vaccines, with 671 of these being original research reports. By setting a threshold of at least 30 citations received before 2020, 272 papers published between 2005 and 2022 were examined. The finding that nearly half of the papers included were published in scientific journals between 2020 and 2022 is explained by the fact that these works received a significant number of citations as preprints or prepublications. We found that 28 papers from this bibliographic portfolio exhibited a Beauty Coefficient following the "Sleeping Beauty" bibliometric phenomenon. Our findings reveal that disruptive technological innovations may be built upon previously neglected reports that experienced sharp citation increases, driven by their crucial applicability to worldwide distresses.

**Keywords:** Backward reference bibliographic search, citation counts, preprints, prepublications, disruptive technological innovation, scientific impact, bibliometrics

## Introduction

In the early 21st century, coronavirus outbreaks caused major emergencies and a global pandemic involving Severe Acute Respiratory Syndrome (SARS-CoV-1), Middle East Respiratory Syndrome (MERS-CoV), and COVID-19 (SARS-CoV-2) (Drosten et al., 2003; Ksiazek et al., 2003; Zaki et al., 2012; da Costa et al., 2020). Despite the first two outbreaks, their limited geographic spread, low number of cases, and lack of vaccines meant the global population remained immunologically unprotected against these pathogens. The emergence of SARS-CoV-2, a new pathogen causing severe pneumonia with high contagion and mortality risks, and no effective antiviral treatments, caught the world off guard and unprepared (da Costa et al., 2020). Governments worldwide, guided by epidemiologists, implemented emergency measures such as social distancing while reallocating funding, resources, and personnel to slow the rise of active cases and expand healthcare capacity (Powell et al., 2021). The focus of all clinical and biological research was to find treatments and develop vaccines against COVID-19 to achieve population immunity (Clemente-Suarez et al., 2020).

The mRNA vaccine technology has been developed for at least three decades, prior to the COVID-19 pandemic, and has been proposed for use against human cancers or pathogens like Ebola or HIV, whereas other methodologies of vaccine development have proven insufficient. However, most of the research focused on animal models, while human testing has mainly remained preclinical. Additionally, concerns were raised about the stability of the mRNA vaccine, its delivery to targeted cells, low efficacy, and excessive immune responses, which limited interest in this technology (Chaudhary et al., 2021). During the COVID-19 pandemic, a new technology of delivering RNA to cells was available, the lipid nanoparticles (LNPs), while previous research on SARS-CoV and MERS-CoV suggested the spike protein as a promising immunogenic target. Capitalizing upon the prior research of small interfering RNA (siRNA) and mRNA vaccines, researchers aimed towards the development of anti-SARS-CoV-2 mRNA vaccines, which exhibited the advantage of rapid, large-scale production to vaccinate the entire population. Still, challenges included the product's stability due to vulnerability to RNases at room temperature, efficient delivery into cells, transportation and distribution difficulties, dosage, and, importantly, the fact that this technology had not been extensively tested on a large scale before (Mulligan et al., 2020). The renewed interest in the clinical use of mRNA vaccine technology has brought previous research by dedicated scientists and persistent companies to the attention of the scientific community.

In a short communication in 2004, Anthony F. J. van Raan proposed an interesting metaphor of the "Sleeping Beauty" in bibliometrics referred in a paper that after sometime of receiving few or no citations, suddenly attracts attention and exhibits a sharp increase in citations received, because of an article mention, the "Sweet Prince" (van Raan, 2004). More than a decade later, Ke et al. (Ke et al., 2015), introduced the Beauty Coefficient of a paper as the calculated value of the maximum citations received per year during a multiyear observation period compared to its citation history, with the citations received on the years of its publication as a reference line. This bibliometric methodology provides the means to detect "Sleeping Beauty" phenomena within a given bibliographic dataset and chronological period of interest.

It could be postulated that the articles about anti-SARS-CoV-2 mRNA vaccines, published during the first year of the COVID-19 pandemic, may serve as "Sweet Princes" for previous papers about mRNA vaccines that are considered as "Sleeping Beauties". Previous reports, based on a direct literature search and beauty coefficient calculation, have shown that the COVID-19 pandemic literature triggered the awakening of prior papers, most of which were about previous coronavirus research (Fazeli-Varzaneh et al., 2021; Haghani & Varamini, 2021; Turki et al., 2022).

Here, we focus on mRNA vaccine research; however, to investigate "Sleeping Beauties," we performed a backward reference bibliographic search, starting from articles on mRNA vaccine development against COVID-19 published in 2020 that may act as "Sweet Princes," and examined their references. It is expected that among these references, there are "Sleeping Beauties" with citation bursts between 2020 and 2023. Interestingly, this method revealed a subset of papers, initially announced and cited as preprints or prepublications, that draw significant attention from mRNA vaccine developers. To our knowledge, this is the first report of preprints or prepublications that later became published in scholarly journals and display "Sleeping Beauty" characteristics.

## Methods

### Literature search and data sources

The purpose of this study was to identify original research papers on mRNA vaccines published before the World Health Organization declared COVID-19 a global pandemic on March 11, 2020 (Cucinotta & Vanelli, 2020), and to analyze their bibliometric patterns, searching for "Sleeping Beauty" phenomena, as previously defined for papers that experienced a sharp increase in citations after a long period of inactivity (Ke et al., 2015). For this purpose, a backward reference search was conducted, starting with potential "Sweet Princes" articles published in 2020 that discussed the development of mRNA vaccines against COVID-19. This investigation was conducted using the Scopus bibliographic database, a well-established resource in bibliometrics for its extensive coverage of scholarly communications (Aksnes & Sivertsen, 2019; Mongeon & Paul-Hus, 2016). The references in these articles were examined to find earlier papers and co-citations. The citations of these referenced papers over the years following their publication were analyzed, after applying the criterion of receiving at least 30 citations up to 2019, to identify "Sleeping Beauties" based on the beauty score calculation for the years 2020 to 2023, as previously described (Ke et al., 2015).

The advanced search query used for published papers that may act as "Sweet Princes" in Scopus was:

```
ALL ( "vaccine development" ) AND ALL ( "COVID-19" OR "SARS-COV-2" ) AND PUBYEAR = 2020 AND ( LIMIT-TO ( DOCTYPE , "ar" ) )
```

The results were exported in CSV format, including citation information, bibliographical details, abstracts, keywords, funding details, and references. Subsequently, a backward reference search was conducted by reviewing the references of the retrieved results, followed by a search within those results for "mRNA vaccin*" and filtering the articles by document type. Finally, the citations of these papers were exported in CSV format from Scopus for further analysis.

## Calculation of the beauty coefficient

The calculation of beauty scores was performed in MS Excel® worksheets by applying the mathematical formula previously described (Ke et al., 2015), which sums the annual citation rates by comparing changes to the highest citation achievement per year, as follows.

$$B = \sum_{t=0}^{t_m} \frac{\frac{c_{t_m}-c_0}{t_m}t + c_0 - c_t}{\max\{1, c_t\}}$$

Whereas $c_0$ is the number of citations in the publication year of the paper, $c_t$ is the number of citations in year $t$ post publication, $c_{tm}$ is the maximum number of citations received by the article in year $t_m$, and max$\{1, c_t\}$ is the highest number of citations from year $t$ onward.

To characterize a paper as a "Sleeping Beauty", a threshold of ≥ 7 was accepted for the Beauty coefficient, since this score indicates a sudden unexpected increase of citations received, in a non-linear fashion, during the evaluation period, which reflects its bibliographic awakening, as previously suggested (Van Raan, 2004). Papers that exhibited a steady and gradual rise in citations without sudden spikes were classified as consistently influential. Papers that achieved low citations during the period of interest were considered dormant.

The VOSviewer was used to map the relationships between articles and identify clusters of highly correlated publications (Van Eck & Waltman, 2010; Van Eck & Waltman, 2014).

## Results

### Comprehensive search retrieved articles

A total of 914 articles about vaccines and COVID-19 were published in 2020, according to Scopus. There is a strong consensus on the subject areas, with 32% of the reports indexed in Medicine, 15% in Immunology and Microbiology, 14% in Biochemistry, Genetics, and Molecular Biology, 8.5% in Pharmacology, Toxicology, and Pharmaceutics, and 4% in Multidisciplinary fields. However, there is considerable variability in the topics of the 5,173 identified index keywords and the 2,042 author-defined keywords, as shown by the analysis of the Scopus-extracted data using VOSviewer (Figure 1). Only 41 (0.8%) index keywords and 4 (0.2%) author keywords meet the criterion of co-occurrence in 10% of the retrieved articles. This variability may be due to the different aspects and scopes defined by the contributing author teams during the COVID-19 pandemic.

Despite this, bibliographic coupling (Kessler, 1963; Martyn, 1964) showed that there are significant information similarities between these articles, with 50% of them sharing at least 20 units of information in common. Co-citation patterns (Small, 1973) differ from bibliographic coupling and reveal that a threshold of 10 citations, as the minimum number of citations for a reference, is met by 0.1% of the cited references. As of June 2025, 41 (4.5%) out of the 914 articles have not received any citations, while the remaining 873 have collectively received more than 119,005 citations, averaging nearly 136 citations per article over a five-and-a-half-year period, and an h-index of 139.

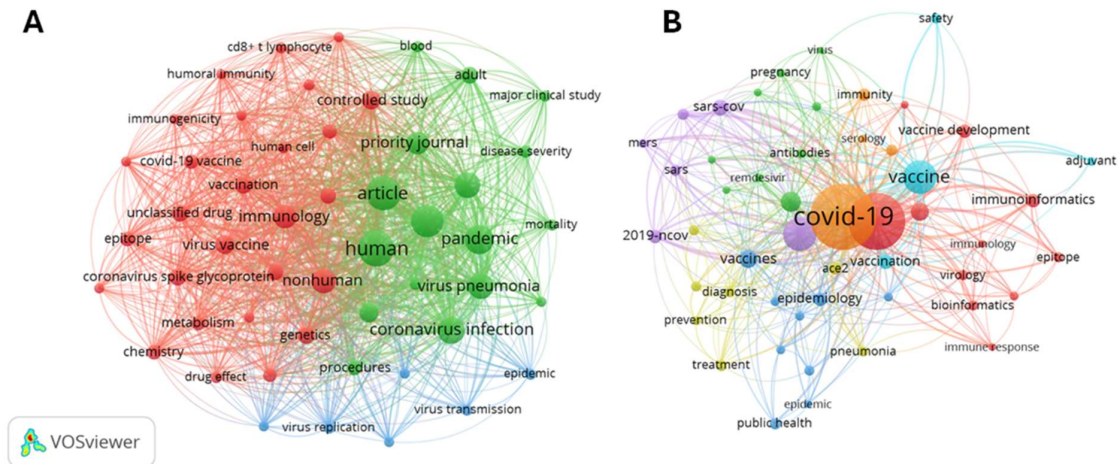

*Figure 1: Co-occurrence map of vaccine-related COVID-19 reports published in 2020 by VOSviewer: (A) index keywords, and (B) author keywords.*

**Backward reference search papers**

Backward reference search, also known as chain search, was considered advantageous over direct keyword or topical search in searching for "Sleeping Beauties" because it allows tracing the citation lineage starting from the awakening citations, the "Sweet Princes," which trigger the spark of recognition of a "Sleeping Beauty" paper. In the case of mRNA vaccine technology, the application-oriented relationship between COVID-19 vaccine-related contributions and the exploration of warp-speed feasible solutions for the pandemic in the literature leads to interdisciplinary reference links. Through this bibliographic investigation approach, the limitations of keyword-based searches were overcome. Backward citation tracking has been proposed as advantageous in scientific and technological domains in identifying dormant inventions (van Raan, 2017).

The total number of references, of the 914 articles about vaccines and COVID-19 published in 2020 and retrieved from Scopus after a comprehensive search, was 21,979 papers. Papers published from 1875 to 2023 have been included in the references, which include a significant body of works that were referenced before their publication as preprints or prepublications. A sharp spiking rise was observed for papers published in 2020, following the eruptive publication record of 87,758 reports and 23,594 preprints issued in the first year of the COVID-19 pandemic (Figure 2). A fraction of 5.4% of them, or 1181 documents, pertained to mRNA vaccines, with 460 out of them to be reviews, 18 book chapters, 17 short surveys, 11 notes, 4 editorials, and

671 original research papers, 5 of which were letters and 5 conference papers (Figure 3). The logistic regression analysis of these papers publication and accumulation trends over time revealed a polynomial correlation between the number of papers and publication years which is significantly compromised by the sharp burst of papers published during the first years of COVID-19 pandemic ($R^2$=0.651), but a more regular exponential correlation between cumulated papers and publications year ($R^2$=0.974) as a result of the same sudden 2020 papers increase. The bibliometric descriptive characteristics of the original research papers on mRNA vaccine technology include a collective of 187,912 citations, averaging 280 citations per paper, with an h-index of 202. Half of these reports concern the subject area of Immunology and Microbiology (52%). In comparison, 37.4% are directly related to Medicine, and applicable clinical research, and 36.7% are pertinent to basic research, Biochemistry, Genetics, and Molecular Biology. The intersection of these three subject areas encompasses reports of translational research, moving from the laboratory bench or animal research to the patient's bedside.

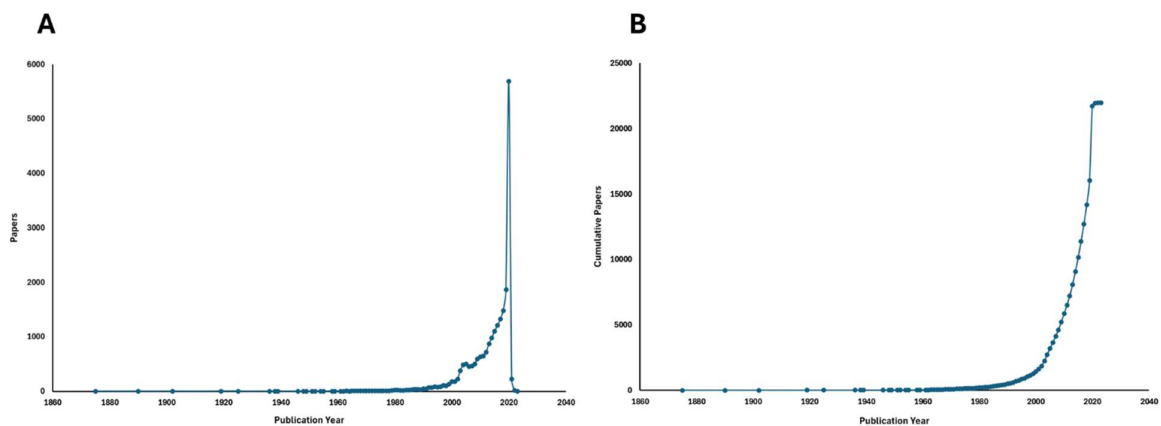

Figure 2: Vaccine-related COVID-19 References distribution over time: (A) Papers per publication year, and (B) Cumulative papers over time.

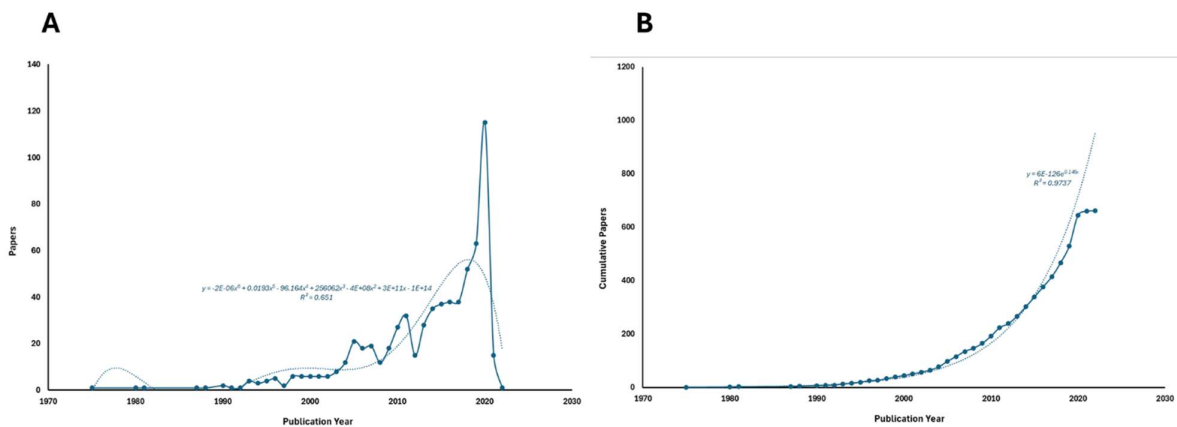

Figure 3: Vaccine-related COVID-19 References on mRNA vaccination(s)/vaccine(s) distribution over time: (A) Papers per publication year (continuous line) with logistic regression polynomial trendline (dotted), accompanied by the predicted equation and $R^2$, and (B) Cumulative papers over time (continuous line) with logistic regression exponential trendline (dotted), accompanied by the predicted equation and $R^2$.

# Beauty coefficient analysis

Citation analysis data were collected from the Scopus database for 671 original research papers on mRNA vaccine technology, identified through a backward reference search of reports related to COVID-19 vaccines. The citation analysis is visually presented as sparklines in a worksheet, as shown for the 48 oldest references (Figure 4). For 272 of the 671 papers, which received at least 30 citations before 2020, the awakening intensity was calculated. This includes the ratio of citations accumulated during the period of interest to the total number of citations received. Additionally, the maximum number of citations per year during the four-year period following the COVID-19 pandemic and the year it began, as well as the highest citations across all years and their respective years, were recorded. The formula for the Beauty Coefficient was applied to these papers as previously described (Ke et al., 2015). The calculations involved creating a pivot table that sets the paper's publication year as 0, then displays subsequent years of the paper's lifetime, along with the individual B ratios from year 0 to year t, which were summed to determine the beauty coefficient for each paper. Twenty-seven papers scored higher than 7 in Beauty Coefficient values, from a minimum of 7.18 to a maximum of 349.50, with an average of 47.79 and a median of 15.72. These papers were published between 2006 and 2020, with an average publication year of 2015.3 and a median of 2016. Representative examples of their citation behavior are depicted in Figure 5. The finding that three papers published in 2020 exhibited characteristics of "Sleeping Beauties" is attributed to the rapid communication scheme applied during the pandemic, as priority prepublications or preprints. All papers recognized as "Sleeping Beauties" with a summary are presented in Table 1.

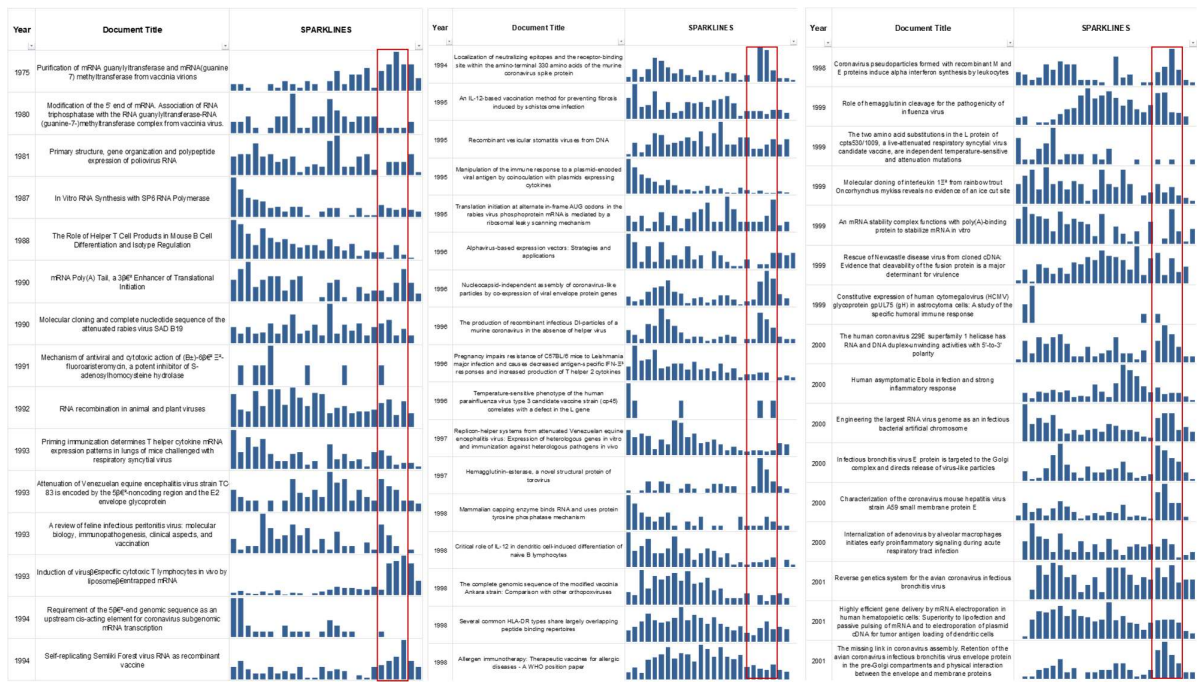

*Figure 4: The 48 oldest references out of 671 original papers related to mRNA vaccine technology, displayed with publication year, title, and citations per year sparklines. The red boxes highlight the four years following the emergence of COVID-19, 2020-2023, which is the time period of interest for identifying "Sleeping Beauties."*

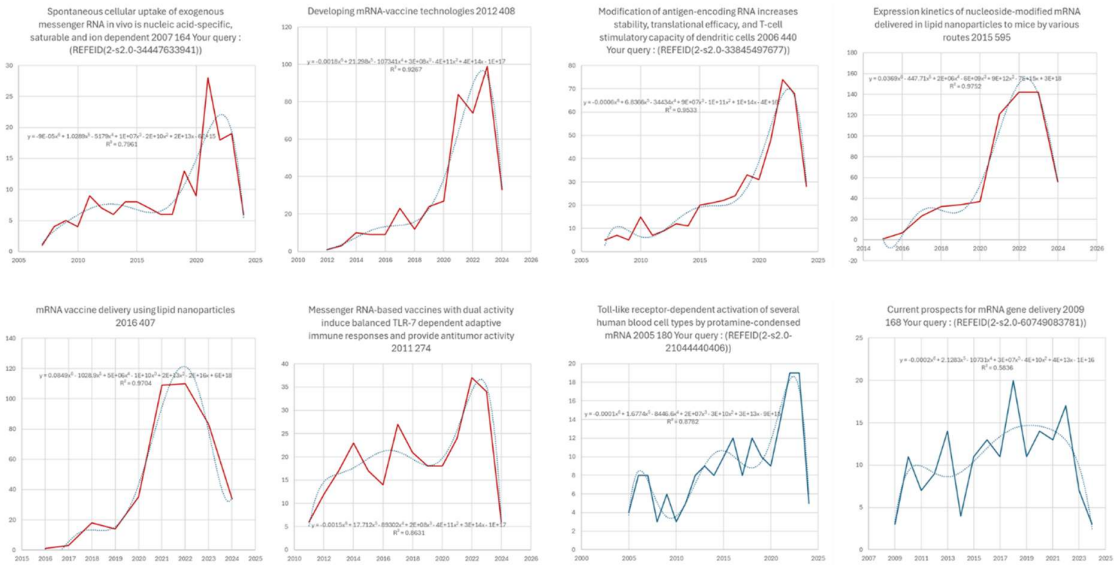

*Figure 5: Representative examples of the citations (y-axis) per year (x-axis) for Sleeping Beauties (red lines) and non-Sleeping Beauties (blue lines). Each graph includes its respective polynomial trendline and R-squared value. Following 2020, citations increased sharply, coinciding with the COVID-19 pandemic and the growing interest in mRNA vaccines within the scientific community.*

Table 1: Sleeping Beauties related to mRNA vaccine technology

| Article Number | Article Title | Year | Authors | DOI | Total Citations | Beauty Score | Summary |
|---|---|---|---|---|---|---|---|
| 1 | Modification of antigen-encoding RNA increases stability, translational efficacy, and T-cell stimulatory capacity of dendritic cells | 2006 | Holtkamp, S., Kreiter, S., Selmi, A., Simon, P., Koslowski, M., Huber, C., TüReci, O., & Sahin, U | https://doi.org/10.1182/blood-2006-04-015024 | 440 | 65,806 | The article examines how modifying antigen-encoding RNA can increase stability, translation efficiency, and the ability of dendritic cells to stimulate T cells, thereby improving the potential of vaccines and immunotherapeutic approaches. |
| 2 | Spontaneous cellular uptake of exogenous messenger RNA in vivo is nucleic acid-specific, saturable and ion dependent | 2007 | Probst, J., Weide, B., Scheel, B., Pichler, B. J., Hoerr, I., Rammensee, H., & Pascolo, S. | https://doi.org/10.1038/sj.gt.3302964 | 164 | 16,676 | The article investigates the natural uptake of exogenous mRNA by cells in vivo, revealing that this process is specific to nucleic acids, saturated, and ion-dependent, thus providing important information for targeted mRNA delivery in gene and immunotherapy treatments. |
| 3 | Messenger RNA-based vaccines with dual activity induce balanced TLR-7 dependent adaptive immune responses and provide antitumor activity | 2010 | Fotin-Mleczek, M., Duchardt, K. M., Lorenz, C., Pfeiffer, R., Ojkić-Zrna, S., Probst, J., & Kallen, K. | https://doi.org/10.1097/cji.0b013e3181f7dbe8 | 274 | 7,807 | The article examines mRNA-based dual-action vaccines, which activate balanced TLR-7-dependent adaptive immune responses, demonstrating their ability to provide anticancer activity through immune stimulation. |
| 4 | Developing mRNA-vaccine technologies | 2012 | Schlake, T., Thess, A., Fotin-Mleczek, M., & Kallen, K. | https://doi.org/10.4161/rna.22269 | 408 | 168,387 | The article presents technological advances in the development of mRNA-based vaccines, focusing on improving their stability, translation efficiency, and immunogenicity, as well as the challenges and prospects for clinical applications. |
| 5 | A novel, disruptive vaccination technology: Self-adjuvanted RNActive ® vaccines | 2013 | Kallen, K., Heidenreich, R., Schnee, M., Petsch, B., Schlake, T., Thess, A., Baumhof, P., Scheel, B., Koch, S. D., & Fotin-Mleczek, M. | https://doi.org/10.4161/hv.25181 | 175 | 7,284 | The article presents an innovative and revolutionary mRNA-based vaccination technology, analyzing its mechanism of action, immune response, and potential clinical applications for the prevention and treatment of various diseases. |
| 6 | Challenges and advances towards the rational design of mRNA vaccines | 2013 | Pollard, C., De Koker, S., Saelens, X., Vanham, G., & Grooten, J. | https://doi.org/10.1016/j.molmed.2013.09.002 | 89 | 17,603 | The article analyzes the challenges and recent advances in the rational design of mRNA vaccines, focusing on optimizing their stability, immunogenicity, and efficacy, with the aim of developing safer and more effective vaccination strategies. |
| 7 | MRNA-based therapeutics-developing a new class of drugs | 2014 | Sahin, U., Karikó, K., & Türeci, Ö. | https://doi.org/10.1038/nrd4278 | 1430 | 93,871 | The article examines the development of mRNA-based therapeutic approaches, presenting the mechanism of action, technological improvements, and clinical applications of this new class of drugs for vaccination, immunotherapy, and gene therapy. |
| 8 | Self-replicating alphavirus RNA vaccines | 2014 | Ljungberg, K., & Liljeström, P. | https://doi.org/10.1586/14760584.2015.965690 | 99 | 78,031 | The article examines self-replicating RNA vaccines based on alpha-viruses, analyzing their mechanism of action, advantages in enhancing the immune response, and potential applications in infectious and cancer immunotherapy. |
| 9 | Lipid-based mRNA vaccine delivery systems | 2015 | Midoux, P., & Pichon, C. | https://doi.org/10.1586/14760584.2015.986104 | 163 | 13,454 | The article examines lipid-based mRNA vaccine delivery systems, analyzing nanoparticle technologies, cellular uptake mechanisms, and challenges in optimizing stability and immunogenicity for more effective vaccine delivery. |
| 10 | A cationic nanoemulsion for the delivery of next-generation RNA vaccines | 2015 | Brito, L. A., Chan, M., Shaw, C. A., Hekele, A., Carsillo, T., Schaefer, M., Archer, J., Seubert, A., Otten, G. R., Beard, C. W., Dey, A. K., Lilja, A., Valiante, N. M., Mason, P. W., Mandl, C. W., Barnett, S. W., Dormitzer, P. R., Ulmer, J. B., Singh, M., . . . Geall, A. J. | https://doi.org/10.1038/mt.2014.133 | 254 | 9,723 | The article presents a cationic nanoemulsion (CNE) as a next-generation carrier for RNA vaccines, analyzing its bioavailability, immunogenicity, and potential for targeted vaccine delivery. |
| 11 | Self-Amplifying mRNA Vaccines | 2015 | Brito, L. A., Kommareddy, S., Maione, D., Uematsu, Y., Giovani, C., Scorza, F. B., Otten, G. R., Yu, D., Mandl, C. W., Mason, P. W., Dormitzer, P. R., Ulmer, J. B., & Geall, A. J. | https://doi.org/10.1016/bs.adgen.2014.10.005. | 129 | 7,805 | The article examines self-amplifying mRNA vaccines, analyzing their mechanism of action, advantages in protein-antigen production, and applications in the development of highly effective vaccines. |
| 12 | Sequence-engineered mRNA Without Chemical Nucleoside Modifications Enables an Effective Protein Therapy in Large Animals | 2015 | Thess, A., Grund, S., Mui, B. L., Hope, M. J., Baumhof, P., Fotin-Mleczek, M., & Schlake, T. | https://doi.org/10.1038/mt.2015.103 | 362 | 10,522 | The article presents a methodology for genetic modification of mRNA sequences without chemical nucleoside modifications, demonstrating its effectiveness in protein therapy in large animal models and its potential for clinical applications. |
| 13 | Expression kinetics of nucleoside-modified mRNA delivered in lipid nanoparticles to mice by various routes | 2015 | Pardi, N., Tuyishime, S., Muramatsu, H., Kariko, K., Mui, B. L., Tam, Y. K., Madden, T. D., Hope, M. J., & Weissman, D. | https://doi.org/10.1016/j.jconrel.2015.08.007 | 595 | 26,973 | The article investigates the expression dynamics of nucleoside-modified mRNA when delivered via lipid nanoparticles (LNPs) in mice, examining the efficacy and differences in efficacy depending on the route of administration. |
| 14 | Particle-mediated Intravenous Delivery of Antigen mRNA Results in Strong Antigen-specific T-cell Responses Despite the Induction of Type I Interferon | 2016 | Broos, K., Van Der Jeught, K., Puttemans, J., Goyvaerts, C., Heirman, C., Dewitte, H., Verbeke, R., Lentacker, I., Thielemans, K., & Breckpot, K. | https://doi.org/10.1038/mtna.2016.38 | 85 | 15,928 | The article examines the administration of mRNA encoding antigens via intravenous particles, demonstrating that it can elicit strong antigen-specific T-cell responses, despite type I interferon induction. |
| 15 | mRNA vaccine delivery using lipid nanoparticles | 2016 | Reichmuth, A. M., Oberli, M. A., Jaklenec, A., Langer, R., & Blankschtein, D. | https://doi.org/10.4155/tde-2016-0006 | 407 | 152,761 | The article examines the administration of mRNA vaccines via lipid nanoparticles (LNPs), analyzing nanotechnological approaches, delivery mechanisms, and challenges in optimizing vaccine efficacy and stability. |

| # | Title | Year | Authors | DOI/Link | Citations | Views |
|---|---|---|---|---|---|---|
| 16 | An mRNA Vaccine Encoding Rabies Virus Glycoprotein Induces Protection against Lethal Infection in Mice and Correlates of Protection in Adult and Newborn Pigs | 2016 | Schnee, M., Vogel, A. B., Voss, D., Petsch, B., Baumhof, P., Kramps, T., & Stitz, L. | e0004746. https://doi.org/10.1371/journal.pntd.0004746 | 154 | 349,502 | The article presents the development and evaluation of an mRNA vaccine encoding the glycoprotein of the rabies virus, demonstrating its protective efficacy against lethal infection in mice and correlates of immune protection in pigs. |
| 17 | Self-amplifying mRNA vaccines expressing multiple conserved influenza antigens confer protection against homologous and heterosubtypic viral challenge | 2016 | Magini, D., Giovani, C., Mangiavacchi, S., Maccari, S., Cecchi, R., Ulmer, J. B., De Gregorio, E., Geall, A. J., Brazzoli, M., & Bertholet, S. | https://doi.org/10.1371/journal.pone.0161193 | 105 | 7,181 | The article explores the development of self-amplifying mRNA vaccines that express multiple conserved influenza antigens, demonstrating their protective effect against homologous and heterologous viral challenges. |
| 18 | Nanotechnologies in delivery of mRNA therapeutics using nonviral vector-based delivery systems | 2017 | Guan, S., & Rosenecker, J. | https://doi.org/10.1038/gt.2017.5 | 267 | 8,659 | The article examines nanotechnological approaches to therapeutic mRNA delivery, focusing on non-viral delivery systems such as lipid nanoparticles and polymers, and the challenges in optimizing their efficacy and safety. |
| 19 | Preclinical and Clinical Demonstration of Immunogenicity by mRNA Vaccines against H10N8 and H7N9 Influenza Viruses | 2017 | Bahl, K., Senn, J. J., Yuzhakov, O., Bulychev, A., Brito, L. A., Hassett, K. J., Laska, M. E., Smith, M., Almarsson, Ö., Thompson, J., Ribeiro, A., Watson, M., Zaks, T., & Ciaramella, G. | https://doi.org/10.1016/j.ymthe.2017.03.035 | 458 | 7,647 | The article presents preclinical and clinical immunogenicity data for mRNA vaccines against H10N8 and H7N9 influenza viruses, demonstrating their safety, induction of strong immune responses, and potential for pandemic preparedness. |
| 20 | Vero cell technology for rapid development of inactivated whole virus vaccines for emerging viral diseases | 2017 | Barrett, P. N., Terpening, S. J., Snow, D., Cobb, R. R., & Kistner, O. | https://doi.org/10.1080/14760584.2017.1357471 | 60 | 14,475 | The article examines Vero cell technology for the rapid development of inactivated whole virus vaccines, focusing on its usefulness in addressing emerging viral diseases and the potential for rapid vaccine production. |
| 21 | Tools for translation: Non-viral materials for therapeutic mRNA delivery | 2017 | Hajj, K. A., & Whitehead, K. A. | https://doi.org/10.1038/natrevmats.2017.56 | 501 | 16,939 | The article examines non-viral methods of therapeutic mRNA delivery, analyzing nanoparticle, lipid, and polymer systems, as well as their challenges and prospects for the effective and safe administration of mRNA therapies. |
| 22 | Safety and immunogenicity of a mRNA rabies vaccine in healthy adults: an open-label, non-randomised, prospective, first-in-human phase 1 clinical trial | 2017 | Alberer, M., Gnad-Vogt, U., Hong, H. S., Mehr, K. T., Backert, L., Finak, G., Gottardo, R., Bica, M. A., Garofano, A., Koch, S. D., Fotin-Mleczek, M., Hoerr, I., Clemens, R., & Von Sonnenburg, F. | https://doi.org/10.1016/s0140-6736(17)31665-3 | 347 | 7,835 | The article presents the results of a Phase 1 clinical study evaluating the safety and immunogenicity of an mRNA vaccine against rabies in healthy adults, demonstrating its ability to induce a strong immune response with a favorable safety profile. |
| 23 | Three decades of messenger RNA vaccine development | 2019 | Verbeke, R., Lentacker, I., De Smedt, S. C., & Dewitte, H. | https://doi.org/10.1016/j.nantod.2019.100766 | 180 | 12,785 | The article reviews thirty years of mRNA vaccine development, examining the technological advances, challenges, delivery strategies, and clinical applications that have shaped the evolution of this innovative vaccination platform. |

| # | Title | Year | Authors | DOI | Citations | Reads | Description |
|---|---|---|---|---|---|---|---|
| 24 | A review of fish vaccine development strategies: Conventional methods and modern biotechnological approaches | 2019 | Ma, J., Bruce, T. J., Jones, E. M., & Cain, K. D. | https://doi.org/10.3390/microorganisms7110569 | 205 | 15,506 | The article reviews vaccine development strategies for fish, comparing traditional methods and modern biotechnological approaches, focusing on improving efficacy, safety, and sustainability in the aquaculture sector. |
| 25 | In silico Design of an Epitope-Based Vaccine Ensemble for Chagas Disease | 2019 | Michel-Todó, L., Reche, P. A., Bigey, P., Pinazo, M., Gascón, J., & Alonso-Padilla, J. | https://doi.org/10.3389/fimmu.2019.02698 | 38 | 12,331 | The article presents the in silico design of a vaccine based on antigens for Chagas disease, using computational methods to select antigenic peptides that can induce an effective immune response. |
| 26 | Lipid nanoparticle technology for therapeutic gene regulation in the liver | 2020 | Witzigmann, D., Kulkarni, J. A., Leung, J., Chen, S., Cullis, P. R., & Van Der Meel, R. | https://doi.org/10.1016/j.addr.2020.06.026 | 197 | 27,767 | The article examines lipid nanoparticle (LNP) technology for regulatory gene therapy in the liver, analyzing transport mechanisms, pharmacokinetic properties, and therapeutic applications for liver diseases. |
| 27 | An evidence based perspective on mRNA-SARScov-2 vaccine development | 2020 | Wang, F., Kream, R. M., & Stefano, G. B | https://doi.org/10.12659/msm.924700 | 179 | 16,924 | The article provides a documented analysis of the development of mRNA vaccines against SARS-CoV-2, examining the technological approaches, efficacy, safety, and challenges encountered during the COVID-19 pandemic. |
| 28 | Opportunities and challenges in the delivery of mrna-based vaccines | 2020 | Wadhwa, A., Aljabbari, A., Lokras, A., Foged, C., & Thakur, A. | https://doi.org/10.3390/pharmaceutics12020102 | 328 | 35,607 | The article analyzes the opportunities and challenges in administering mRNA vaccines, examining the delivery technologies, stability, immunogenicity, and pharmacokinetic parameters that affect their efficacy. |

## Discussion

To explore the "Sleeping Beauty" bibliometric phenomenon, the case of mRNA vaccine technology used against COVID-19 was examined. The advancements in siRNA technology in lab research following the discovery of the RNA interference (RNAi) biological mechanism were numerous (Schutze, 2004), but mRNA vaccine research was progressing slowly before the COVID-19 pandemic (Chaudhary et al., 2021). The development and worldwide vaccination of human populations with the mRNA vaccines BNT162b2 by Pfizer-BioNTech (Polack et al., 2020) and mRNA-1273 by Moderna (Baden et al., 2021) represent disruptive applications of an existing, but not widely produced or tested, technology. A retrospective evaluation of previous mRNA vaccine research was necessary to rapidly prepare, combine, and test the compounds needed for this purpose. The development of mRNA vaccines against COVID-19 was not a "miracle," as such a label overlooks the decades of basic research and technological advances that enabled this rapid and successful delivery; it was a scientific breakthrough. The foundational "roots" of mRNA vaccine technology were examined here, including previously overlooked papers that were bibliographically awakened because of the pandemic to be in the spotlight, the "Sleeping Beauties."

The literature search approach employed in this investigation was a backward reference search, as this method has been proposed as advantageous in detecting the course of scientific or technological breakthroughs (van Raan, 2017). This methodology considers the research trajectory following the path of reference to previous papers, whilst bypassing terminology issues that may arise from a keyword search. Starting from a well-defined set of articles that discuss and analyze the available opportunities for anti-SARS-CoV-2 vaccines early in the pandemic, it was possible to expand the literature pool examined by tracing their references backward. In this context, by applying the previously described Beauty coefficient (Ke et al., 2015), it was possible to identify "Sleeping Beauties," essential papers on mRNA vaccine technology, published more than a decade before the pandemic. What was surprising is the identification of papers published in the same year with the potential "Sweet Princes," that perfectly align with the "Sleeping Beauty" definition. These papers represent the vigorous and intense bibliographic surge of the COVID-19 literature, and the emergency actions by the scientific publishers, introducing fast-track review, open reviews and prepublication announcements without paywall restrictions, as well as the increase in the use of preprints as a valid form of scientific communication (Chaleplioglou & Koulouris, 2023).

The value of this analysis is further emphasized when the findings on the awakening of scientific publications are linked with previous bibliometric studies that examined phenomena of delayed recognition or unexpected increases in influence in the field of mRNA vaccines. Although several studies have focused on the overall research trends of mRNA technologies (Wang et al., 2020; Liu et al., 2022), few have systematically analyzed the mechanism of bibliometric awakening, which refers to the sudden elevation in citations to older articles that had remained relatively unnoticed for years. A comparison with the study by Ye et al. (2021), which examines the resurgence of scientific articles in medicine and biotechnology, reveals that the pandemic catalyzed the uncovering of previously overlooked knowledge. Chen et al. (2021) observed that the most frequently cited articles during the pandemic were not necessarily recent, but rather those offering basic methodological or technological solutions that could be found in the existing body of scientific knowledge. This finding is supported by the current analysis, which shows that mRNA vaccine-related studies from the first decade of the 21st century exhibit initially few citations and surpassed the Beauty coefficient value of 7 threshold after 2020, making them typical "Sleeping Beauties." Notably, previous investigations utilizing the Beauty coefficient showed that COVID-19 literature triggered the awakening of past coronavirus research papers (Fazeli-Varzaneh et al., 2021; Haghani & Varamini, 2021; Turki et al., 2022).

One aspect that distinguishes this study from previous ones is its starting point, the potential "Sweet Princes," and their logical interpretation of these articles in relation to the awakened "Sleeping Beauties." The relationship between these papers closely aligns with co-citation bibliometric patterns. However, it should be clarified that the identified "Sleeping Beauties" were not the sole driving forces behind COVID-19 mRNA vaccine research. The influence of consistently highly cited papers should not be overlooked or underestimated. The awakening of "Sleeping Beauties" represents a bibliometric phenomenon significantly amplified by the substantial bibliographic output in the first four years following the emergence of COVID-19. Additionally, beyond the mRNA vaccine, other topics were also important and interesting, including hospitalization and management of COVID-19 patients, epidemiological measures to reduce contagion, research on pharmacotherapeutic agents, and psychological and socioeconomic issues resulting from social distancing and lockdowns. Some of these topics may also have "Sleeping Beauty" phenomena related to COVID-19. Other limitations of this study include its focus on bibliographic data rather than practical development of mRNA vaccines, reliance on a single bibliographic database instead of multiple sources, strict inclusion criteria for papers in the analysis, and the tight threshold of the beauty coefficient used, which may exclude some contributions. Nonetheless, it is important to note that successfully aligning bibliometric data from multiple sources has not been reported, given the inherent differences among databases, including coverage, indexing, and journal inclusion and exclusion criteria (Aksnes & Sivertsen, 2019; Mongeon & Paul-Hus, 2016). Variations observed in logistic regression indicate specific characteristics of some papers, such as the possibility of gaining attention before 2020 or suddenly attracting extensive attention and many citations after 2020. The logistic regression coefficients reflect the goodness of fit of the model as a predictor of citation trends driven by the data, without considering broader effects of scientific or technological growth over time.

## Conclusions

Backward reference search, starting from articles that may serve as "Sweet Princes" to bibliographically awake past papers, and following their references to identify "Sleeping Beauties," is an inclusive and targeted approach that overcomes the limitations of keyword-based searches. The mRNA vaccine technology, which was advancing slowly and with related papers receiving limited recognition in terms of citations before 2020, became the focus of research funding and efforts within the first months of the COVID-19 pandemic. Awakened "Sleeping Beauties" were identified in the corpus of mRNA vaccine technology. A subset of reference papers of interest are those published during or after 2020 that existed as prepublications or preprints earlier and received citations. During the COVID-19 vaccine bibliographic surge, these reports gained increased attention, exhibiting "Sleeping Beauty" characteristics. Disruptive technological advances may be driven by necessity and bring previously neglected reports to scientific attention.